\documentclass[pre,floatfix]{revtex4-1} 
\usepackage[pdfpagemode=UseNone,pdfstartview=FitH,colorlinks=true]{hyperref}
\usepackage{enumitem}
\setlistdepth{9}
\usepackage{graphicx}
\usepackage{epstopdf}

\begin{document}
\title{Collective regulation by non-coding RNA}

\author{J. M. Deutsch}
\email{josh@ucsc.edu}
\affiliation{Department of Physics, University of California, Santa Cruz CA 95064}

\begin{abstract}
We study genetic networks that produce many
species of non-coding RNA molecules that are present at
a moderate density, as typically exists in the cell. The associations
of the many species of these RNA are modeled physically,
taking into account the equilibrium constants between bound and unbound states.
By including the pair-wise binding of the many RNA species, 
the network becomes highly interconnected and shows different
properties than the usual type of genetic network.
It shows much more robustness to mutation, and also rapid evolutionary
adaptation in an environment that oscillates in time. This provides
a possible explanation for the weak evolutionary constraints seen
in much of the non-coding RNA that has been studied.
\end{abstract}
\maketitle
 
\section{Introduction}

Recently there has been a great deal of interest in the function of non-coding
RNA molecules (ncRNAs) that were previously thought to be mostly
nonfunctional~\cite{ENCODEpilot,Kapranov2007,MercerDingerMattick,ENCODE}. Most mammalian genomic transcripts do not directly
code for proteins and only approximately 5\% of the bases can be confidently identified 
as being under evolutionary constraint. Many functional elements appear
to have little constraint and several hypotheses have been advanced
to explain this~\cite{ENCODEpilot}, but no clear explanation has emerged. 

Here we examine the physical constraints on ncRNA based on their high
density and the general level of binding between molecules. This leads
to the possibility of regulation due to collections of many different sequences of RNA.
Each species by itself has only a small effect on function, but taken
together as a group, this pool of molecules can ``collectively regulate"
a network of genes. By having a large pool of species bind to each other
and mRNA with different
affinities, the unbound RNA will decrease by an amount that depends
on the concentrations of all of the other types of RNA, not just a few species. 
We will demonstrate that this kind of regulation is far more robust against mutations than
the standard means for regulating gene expression which involve far fewer interactions.
Moreover networks of this
type evolving under environments that are oscillatory in time will evolve to
a state where the ncRNA binding interactions encode a memory of its past. As a result,
when put
in a previous environment, the population of networks will quickly adapt to it. This
provides an explanation for functional elements that individually appear to
have little evolutionary constraint.

\section{Is ncRNA highly interactive?}

Discussion of hybridization of different RNA molecules is normally in a context
where potential interaction between them is assumed to be possible. This presupposes
that the concentration of such molecules is sufficient so that interaction between them
will occur on a biologically relevant timescale. It is therefore worthwhile understanding
(a) how long it takes for two ncRNA molecules to collide and
(b) if two such arbitrary ncRNA molecules are likely to show significant hybridization.

The overwhelming majority of transcription taking place in human cells is not associated
with protein coding genes, but produces long non-coding RNA. Estimates of total
RNA concentration in cells range by an order of magnitude but are typically of
order $10^3 ng/\mu l$, which for an RNA molecule of 100 nucleotides (nt) corresponds to an average separation of $l = 33 nm$. 
We can estimate the collision time scale for a molecule of $N$nt
to collide with another of the same approximate size. The diffusion coefficient of a 20nt single
stranded RNA is approximately $5 \times 10^{−6} cm^2/s$~\cite{Yeh}, and conservatively 100nt is 1/5 of
this value, $D \approx 10^{-6}cm^2/s$. This means that that average time to a collision is 
$t \approx l^2/(2D)(l/R)$. The factor of $l/R$ is the random walk collision probability to an object
of average size $R$~\cite{BergRandomWalksinBiology}. The value of $R$ is approximately $20 A$~\cite{Hyeon}, giving
$t = 10^{-4}  s$. 
Therefore we expect there to be many collisions between RNA molecules
on the timescale of seconds, which is still less than the relevant timescale for transcription.

Now we ask what happens if two arbitrary ncRNA collide. Because of the need for specificity
in many situations in biology, it would seem likely that such collisions would mostly be
inert, with little binding between the molecules. However experimental evidence
and theoretical analysis makes this situation far less clear.
For example, it is evident from accurate modelling~\cite{Zuker,Mathews} that even randomly 
sequenced RNA molecules are expected to show a large degree of internal secondary structure.
It has been shown that ~\cite{WorkmanKrogh,Clote} that the free energy
of structural RNA sequences is only modestly lower than in random sequences with 
the same nucleotide frequencies. The potentials that were used in this work,
are predominantly attractive even for mismatched base pairing. By forming loops
and hairpins, the potential can be further lowered, increasing the binding affinity.

Therefore, we must take seriously the possibility that
there are many interactions between arbitrary ncRNA molecules.
Clearly there will be a distribution
of interaction strengths, were some are tightly hybridized, in the case of well
matched sequences, whereas others will only hybridize weakly.
Without a clear experimental reason why such interactions are not present
for ncRNA, it behooves us to consider their effect. We will now explore the consequences of having many significant
interactions between ncRNA.

\section{Collective Interactions Modify RNA Levels}
\label{sec:CollectiveInteractions}

NcRNA has been shown to effect the transcription of RNA by many mechanisms, and
therefore a modification of the concentration of unbound ncRNA should affect transcription.
We can consider a number of different species of RNA interacting with each other
through partial hybridization. We will assume that these interactions are in fact
weak, and that one RNA is capable of interacting with many other species,
because the RNA are capable of weak binding to different sites along the same
molecule. 

Let us focus on one species, $j$, with a total concentration of $C_j$, some of which, $\rho_j$,
will be completely unbound to other RNAs. Assume there are a total of $N$ species 
of all RNA's. 
We model this as a set of chemical reactions of the form $i + j \rightleftharpoons ij$.
Writing the of bound pairs $ij$, we have the equilibrium constant~\cite{Reif} $K_{i,j} = \rho_{ij}/\rho_i\rho_j$.
Therefore the total amount of $j$ bound to some other molecule is
\begin{equation}
\label{eq:B=}
C_j-\rho_j = \sum_i \rho_{ij} = \sum_i K_{i,j} \rho_i\rho_j
\end{equation}
and rearranging this we obtain
\begin{equation}
\label{eq:ConcentrationModification}
\rho_j =\frac{C_j}{1+\sum_i\rho_i K_{i,j}}
\end{equation}
Although the model assumes only a collection of bound to unbound reactions,
the final result is intuitively reasonable: the amount of unbound RNA
decreases with the total amount of other RNA molecules interacting with it,
weighted by the binding affinity for each species to the one of interest.
We could of course use a more specific and realistic model of this equilibrium,
but that would not be useful to the general points that we are making here.

\section{Modelling genetic networks}

The ultimate purpose of regulatory networks is to help in performing biological functions.
There will be inputs to it, which ultimately represent external factors,
for example, temperature or lactose concentration. Without losing any mathematical
generality, these can also be considered to be protein concentrations acting on
input nodes of the network. After being processed by the network, the outputs will be the levels of
gene products that are biologically relevant, for example heat shock proteins or $\beta$-galactosidase. 
An example of such a network is shown in Fig. \ref{fig:network}(a).
To fulfill a useful function, the outputs that are generated will  depend on the combination of inputs
presented to the network. We can quantify this input-output relation
by quantizing the levels to two values, $0$ and $1$ and regard this as a general
digital circuit. The circuit
performs the task of mapping $N_i$ input values, to $N_o$ output values. 
For example it might map $1,0,1 \rightarrow 1,0$, and  $0,0,1 \rightarrow 0,1$.
The complete behavior of this quantized map is characterized by knowing how
all $2^{N_i}$ inputs map to outputs. The number of such distinct input-output
relations grows very rapidly with $N_i$ and $N_o$.

To understand the possible role of collective ncRNA effects, we will employ
evolutionary algorithms for two reasons. First, to try to understand the role of
such collective effects in enhancing
an organism's ability to adapt, through mutation, to a changing environment. 
Second, we want to design a circuit that gives the correct
outputs to the network's inputs.  
We start in section \ref{subsec:NetworkWOncRNA} by designing networks without 
ncRNA. Then in section \ref{subsec:NetworkWithncRNA}, we show how the ncRNA can be modeled to modify the network
architecture to include collective regulatory effects. The effects of the ncRNA
on network behavior are then explored.

\subsection{Network architecture without ncRNA}
\label{subsec:NetworkWOncRNA}

In order to be able to include the molecular interactions discussed above, 
we use  models that are continuous in the concentrations of the constituent components.

We first consider a genetic network in the absence of ncRNA using a model of 
{\em cis}-regulation with two types of regulatory elements, enhancers and silencers. 
Each gene is regulated by a number of such
elements. Together, these control the expression of a gene protein
which in turn can act to further regulate other elements of this network.

\begin{figure}[htb]
\begin{center}
\includegraphics[width=0.7\hsize]{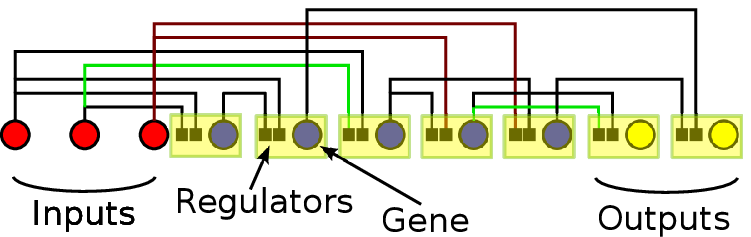}\\
(a)\\ 
\vskip .1in
\includegraphics[width=0.7\hsize]{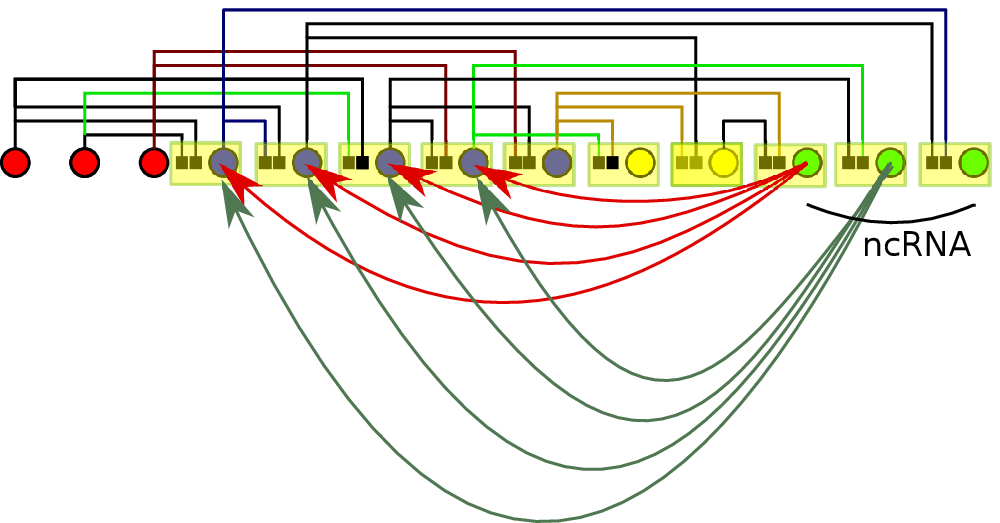}\\
(b) 
\caption
{ 
(a) A {\em cis}-genetic network. Input protein concentrations act to regulate
intermediate gene levels, which further regulate other genes. Regulatory
elements (squares) act upstream of a gene (circle), and these can act to enhance
or to silence a gene. After processing through intermediate genes, final output
genes expression values can be produced in response to the network's inputs.
(b) The addition of ncRNA to the network. The inputs are on the left, and the
outputs are still colored yellow.  The DNA coding for the ncRNA is
regulated by a combination of enhancement and suppression, similar to the genes
in (a). The ncRNA interact with all genes through collective binding as
described in section \ref{sec:CollectiveInteractions}
}
\label{fig:network}
\end{center}
\end{figure}

For enhancement of a gene expression level, the binding of individual enhancer regulatory proteins 
at concentration $c$, will 
change expression by a factor $l_e(c)$. We assume that $l_e$ is a nonlinear function
of concentration, initially increasing slowly until some threshold and then
rising sharply before leveling off. Similarly, a silencer will have the reverse
effect, suppressing expression by a factor $l_s(c)$. In certain cases, these functions
have been measured in detail~\cite{Setty}. For $l_e$ we choose a sigmoidal shape,
\begin{equation}
\label{eq:logit}
l_e(c) = \frac{1}{1 + \exp(-A(c-c_0))}
\end{equation}
although many
functions with the same asymptotic behavior are expected to behave similarly.
Suppression of expression will appear to be the mirror image of this, $l_s(c) = 1-l_e(c)$.
The final value of the expression level is assumed to be proportional to the product of all $m$
such factors, each one of which is either an enhancer or silencer. Every gene has
a regulatory module that determines the level of gene expression $E$,
\begin{equation}
\label{eq:productoflevels}
E = \prod_{i=1}^m l_{\alpha_i}(c_i)
\end{equation}
where $\alpha_i$ can be either an enhancer $e$, or silencer $s$, and $c_i$ is
concentration of the specific protein that binds to this element.

This network can be readily modelled numerically.
The input levels are fixed, and the gene concentrations are calculated according
to Eqs.  \ref{eq:logit}
and \ref{eq:productoflevels}. This process is iterated until steady state values
are reached, or terminated if a limit cycle is detected, or convergence is not
reached after a maximum number of iterations (normally set to $20$). To prevent
limit cycles, we use feed-forward architecture, where the output must feed into
a gene to its right in Fig. \ref{fig:network}(a).

\subsection{Addition of ncRNA}
\label{subsec:NetworkWithncRNA}

Although the range of mechanisms and functions of ncRNA is large and still
under much active investigation, it is already apparent that it can act
in a manner similar to a regulatory gene.
NcRNA transcription is known to be regulated by a variety of mechanisms
including epigenetic modification and transcription
factors~\cite{Khalil,Mohamed,Guttman}. ncRNA will
modify transcription of genes and can act as a silencer or an enhancer~\cite{Orom}.
Therefore when including the effects of ncRNA, the general architecture of the last section will still remain
valid with the reinterpretation that some genes can now be ncRNA whose
concentration will affect the expression of other genes. We will call
such units ``generalized genes". This is in contrast to
a ``coding gene", where the production of a protein, not only RNA, also occurs.

The fact that RNA can take part in regulation means that we should
include such generalized genes in order to get a complete description of all of the architectures
possible. We will focus on adding in extra elements that do
what we termed ``collective regulation". We consider the expression
of ncRNA on the right in Fig. \ref{fig:network}(b). These ncRNAs
play  the role of the ensemble of molecules discussed in section
\ref{sec:CollectiveInteractions}. They will modify the RNA
concentrations produced by the generalized genes according to Eq. \ref{eq:ConcentrationModification}.
The presence of this large collection of ncRNA will affect the output
concentrations, still shown in yellow.

To quantify these effects, we will consider networks where all the generalized
genes are influenced by the ncRNA. Doing this is the best way to illustrate the
collective effects that are being proposed, but this simplification is not
expected to alter the general phenomenon. Eq. \ref{eq:productoflevels} is modified by
multiplying it by Eq. \ref{eq:ConcentrationModification} that diminishes
the expression of a generalized gene due to promiscuous binding with ncRNA.

Therefore the algorithm iterates the following procedure, keeping track of
both the total current levels of expression, and the free levels that remains
after suppression:
\begin{itemize}
\item {\bf Collective suppression.} Given the current level of expression of the generalized genes,
calculate how this is suppressed by Eq. \ref{eq:ConcentrationModification}.
\item {\bf Regulation of expression.} Consider each gene's promoters. Use 
Eq. \ref{eq:productoflevels} to calculate the expression level of each generalized gene, 
equating the suppressed values calculated in the above step with the $c_i$'s
\end{itemize}
until constant expression levels are reached, or a limit cycle is detected, as
described above in section \ref{subsec:NetworkWOncRNA}.

The number of independenet rate constants is determined by
the physical restriction that $K_{i,j} = K_{j,i}$ and $K_{i,i} = 0$.  
With $N_{nc}$ types of ncRNA present, and $N_c$ coding genes (excluding inputs), this means
there are $(N_{nc})(N_{nc}-1)/2 + N_{nc}N_c$ 
couplings of non-coding to non-coding and non-coding to coding elements.

\subsection{Evolutionary Algorithm}

The goal for a network is to learn a specific rule, which maps all $2^{N_i}$ 
possible inputs to outputs. The network will start off making many
errors but will evolve so as to minimize these.
We will start with a population of networks $N_p$ and for each one, measure 
the fraction of mistakes that are made. That is, for all $2^{N_i}$ inputs
we compute the
difference between how the network maps the inputs to the outputs, and the specific
rule that we want it to achieve. The fitness used to decide how to replicate
systems, is almost the same. The outputs of the simulation are analog, with
concentrations varying continuously between $0$ and $1$. The magnitude of the difference between
that analog output and the binary output of our goal, is used as a measure of the fitness. The
reason that this measure is chosen rather than the number of mistakes, is that
information is lost in doing this binning and we
should select for networks that are closest to the binary goal.

We randomly mutate each network in the population, changing the connectivity
and adding and reducing promoter elements.  However we keep the number
of generalized genes and ncRNA constant. The number of times
a network is replicated or eliminated, depends on the degree to which the
mutations decrease the mistakes it makes. This is done through an evolutionary replication
algorithm that has been used frequently in many contexts~\cite{Orland}.

\begin{figure}[htb]
\begin{center}
\includegraphics[width=0.4\hsize]{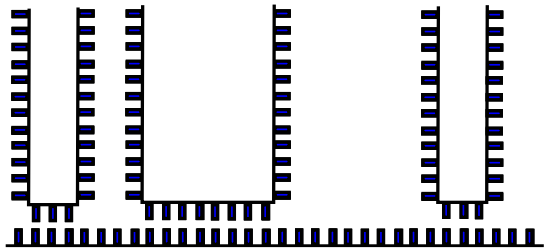}
\caption
{ 
The hybridization of an RNA molecules to a choice of several others. The
sequences have been designed so that the regions of complementarity, and
therefore binding, occurs in non-overlapping regions. This means that the binding
energies can be designed to be nearly independent of each other.
}
\label{fig:hybridization}
\end{center}
\end{figure}

In addition to changing promoter properties, we change the values
the binding free energies and hence rate constants, $K_{i,j}$ for the ncRNA. We are making
the assumption that it is possible to choose any set of binding
energies that we please by adjusting the RNA sequences appropriately. 
That is, although the binding energies between
any two molecules may affect other interaction strengths, it is always
possible to find a set of RNA sequences with the desired binding strengths.
The error in this approximation is hard to quantify, but for long RNA molecules
it should become possible to find sequences that can hybridize independently,
making it possible to choose binding strengths for each one independently.
See Fig. \ref{fig:hybridization}.

In the following, we restrict the number of promoters, $m_p$, associated with
a gene to  $1 \leq m_p \leq 2$ and set $N_p$ to $16$.
If the number of inputs $N_i = 3$, and the number of outputs $N_o = 2$,
a network can be evolved to correctly learn random rules. This can be
accomplished with having $N_c = 9$ coding genes  (excluding inputs), even
in the absence of ncRNA, that is $N_{nc} = 0$. This was tried with $7$ independent
sets of rules and the system succeeded in learning correctly in all cases.
This usually is accomplished depending on the random rule, in less than $40000$ generations.
If $N_{nc}$ is nonzero,
the network is also able to learn the correct rule. However there is no compelling
argument why it would be 
evolutionary advantageous to include ncRNA in a static environment where a fixed rule is being
learned. We will see later however, that collective interactions are
more robust to mutations in the network. There is however a more clear
reason why ncRNA should be useful in a changing environment.

\subsection{Adaptation to changing environments}

We can ask what happens if the environment of the organism fluctuates
in time, so that the network must periodically re-adapt to changing
conditions. The effects of adaptation on a species' fitness have been
investigated in relation to host-parasite dynamics and in particular the
``Red Queen Hypothesis"~\cite{VanValen}.

If environmental conditions change, for example, the increase in salinity
of the environment, a network can eventually learn to adapt to this
new condition. If the conditions then returns to their original state,
the network needs to relearn the original set of rules. Organisms that
can keep a memory of previous conditions will have an advantage over
those with no memory. We can ask if ncRNA helps in re-adapting to previous
conditions. Intuitively, because the $K_{i,j}$'s give a large number of adjustable
parameters, there is a memory contained in these for previous
adaptations. In other words, if we train the network to learn rule A,
and then change to rule B, the network will still retain some information
about rule A in its ncRNA interactions.

To investigate if adding in collective interactions helps in periodic
adaptation, two separate rules were used. A random rule was generated,
and then a mutation was made so that one of the mappings from one
input, e.g. $1,0,1 \rightarrow 1,0$ was altered so that the output differed
in one position, e.g. $1,0,1 \rightarrow 1,1$. The rule the network is
learning is altered periodically every $N_p$ generations. 

\begin{figure}[htb]
\begin{center}
\includegraphics[width=0.7\hsize]{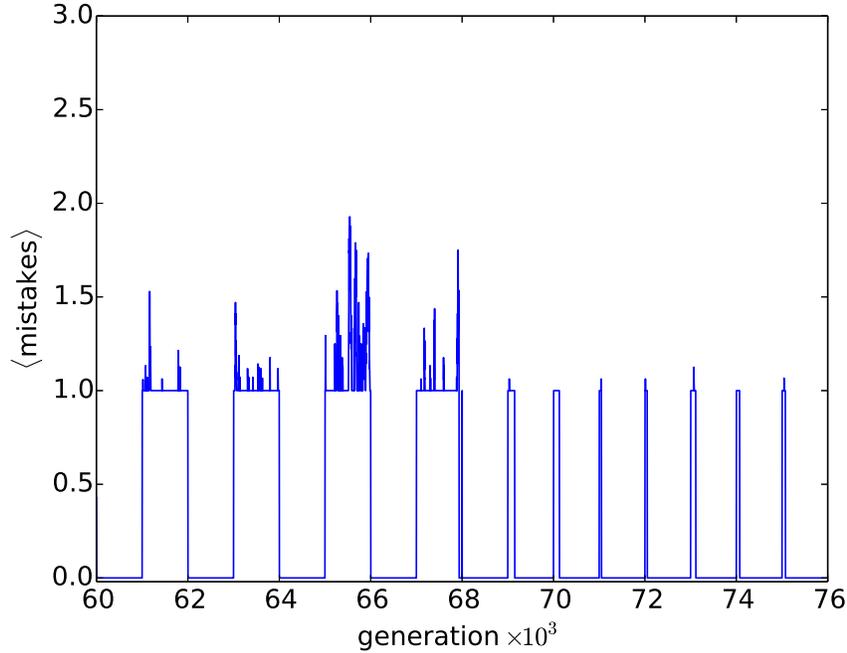}
\caption
{ 
The average number of mistakes made as a function of generation for a population
of $16$ gene networks
which includes the effects of ncRNA collective regulation. Every $500$ generations
the environment switches requiring adaptation to different rules. As the system
evolves, it the system makes fewer mistakes, but as the environment switches,
the number of mistakes goes up. By the end of the run, the population of networks
has evolved to quickly adapt to these changes.
}
\label{fig:mistakes}
\end{center}
\end{figure}

It is not necessarily the case the we can provide the network with
the knowledge of the condition. If the condition is, for example,
the re-emergence of a predator, it is not obvious how this new
condition could be given to the network as an input. Therefore we examine the
case where this input is omitted. This means
that this problem is no longer equivalent to learning a single but
more complex rule. There are now two rules that are
contradictory, and the network must evolve quickly to adapt to
this periodic change.

Because the two rules are the same for all but one input, initially
the population's evolution should be similar to that with a fixed
rule. As the system evolves and makes fewer mistakes, it will get
into a regime where it will not be able to match both rules unless
it adapts quickly enough. 

Even after successful
adaptation has taken place, mutations will always cause some
individual networks to perform non-optimally. Therefore to
obtain a measure for how well the population has
adapted we consider their most fit specimens. That is, we consider the
minimum number of mistakes made by individuals in the population.
Because we are varying the environment between two states, we
can see if adaptation has occured by
examining the number of mistakes right before switching. This 
tells us the fitness in one environment, but to get a measure
of fitness for both environments, we consider the the average of
this minimum in the current phase and in the last one, which we
label $M$. The network
performs perfectly if $M=0$. If it has adapted correctly to
only one environment and not the other, then $M = 1/2$ instead.
However perfect performance can be short lived, so the algorithm
will continue to evolve to check if the population found is stable.
It checks for stability by only terminating the process if
for the last $16$ changes in the rules $M=0$.

The simulation was run with the same parameters as above but with
$6$ coding gene nodes and $6$ ncRNA. The rules, as above, were switched
every $N_p = 1000$ generations. The simulation either terminated due
to the above criterion, our it reached a ceiling of $8\times10^6$ generations.
This simulation was tested with $7$ statistically independent sets of rules.
At the end, the value of $M$ averaged over these $7$ systems was $0.07$. 
In contrast when we perform the above runs with the same conditions but no ncRNA, 
this average was $0.57$ meaning that the systems did not adapt well.

Fig. \ref{fig:mistakes} shows the
average number of mistakes made for the whole population learning two rules starting at $60,000$
generations and running to completion at $76,000$ generations.
The number of mistakes increases after a shift of rules, but then
the population adapts to the new rule quickly enough to keep up
with this change. This is evident from the spikes in mistakes after
a switch which drops much before the next environmental change takes place.

\begin{figure}[htb]
\begin{center}
(a)
\includegraphics[width=0.4\hsize]{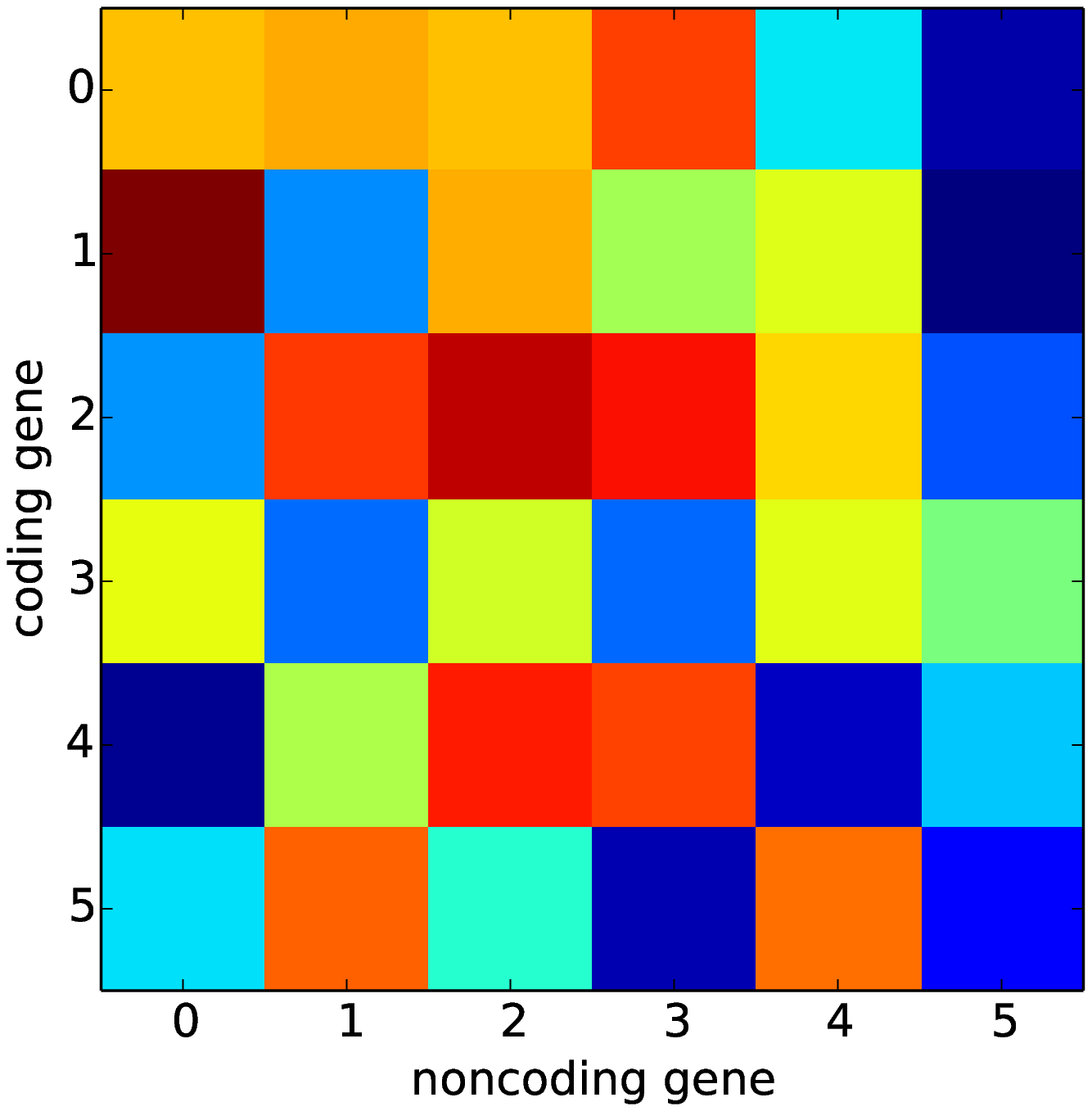}
(b)
\includegraphics[width=0.4\hsize]{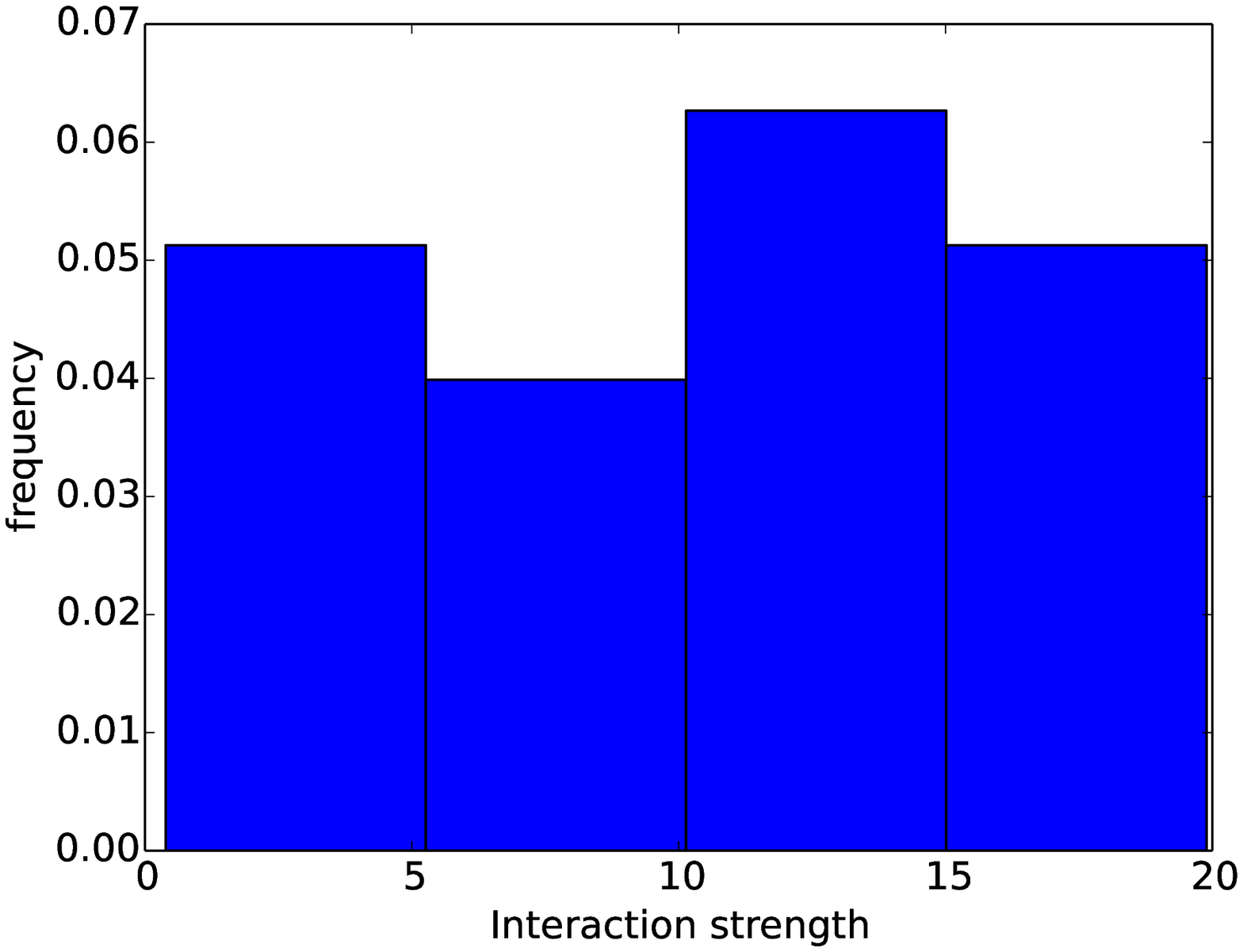}
\caption
{ 
(a)The couplings $K_{i,j}$ between noncoding and coding RNA's that are found after 
a system has evolved to quickly switch between different rules.(b) A histogram
of the same data.
}
\label{fig:interactions}
\end{center}
\end{figure}

It is of interest to examine the distribution of interactions $K_{i,j}$
that are produced by this procedure. Fig \ref{fig:interactions}(a) shows these
interactions with the strength color-coded. Fig \ref{fig:interactions}(b) 
shows a histogram of the same data. Note the broad distribution of interactions.
The interactions are highly non-local, involving coupling of all of the
different RNA molecules.

\subsection{Robustness to changes}

Taking the same system as was analyzed and shown in Figs.  \ref{fig:mistakes} and  
\ref{fig:interactions} we can examine what happens if we randomly delete some
of the interactions, so that some of the $K_{i,j}$ 
are randomly set to zero. With such deletions, the network was tested to
see if it still displays the last rule that it had been evolved to reproduce.
Altogether there are $6\times6 + 6\times(6-1)/2 = 51$ independent $K_{i,j}$.
All seven statistically independent systems displayed a considerable degree of
robustness. With 25 random deletions for each system, they all still
reproduced their rule perfectly. 
In contrast, for $6$ networks that successfully learned one rule, with $7$ coding genes and no 
ncRNAs, a single deletion of a promoter had a significant effect. Only $1$ of
the $6$ networks was still able to correctly reproduce the initial rule.
Therefore the system with non-coding RNA can withstand a high degree of mutation compared
to a more standard network.

\subsection{Discussion}

We have argued that the high concentration of ncRNA in the cell and
the high affinity for binding, suggests that it is necessary to take these
many interactions into account, rather than a subset of specific
ones, as is normally done. If these many interactions are indeed present, then
there is a need for a new way of looking
at gene regulation. Instead of genes being regulated only by a few
specific interactions, they are additionally regulated by a soup
of interacting RNA molecules that collectively form a genetic network
capable of performing biologically useful functions.

It should be noted however, that this model does not assume that all
molecular species bind together at the same time. A given molecule
has a probability of pair-wise binding and unbinding to many different species, 
so that regulation is achieved by affecting the average concentration of free
RNA, and does not directly regulate individual molecules.

The model we employed greatly simplifies the extremely complex
nature of genetic regulation. We are assuming that any two
RNAs will hybridize quite weakly, and in this bound state, they are
inert and not capable of performing any biological function, such
as being transcribed into a protein. This certainly not true. For
example, two long RNA chains can likely bind to a third one. This
is not considered. However it is not hard to add in such interactions
and it is not expected to qualitatively alter our conclusions. We used
a simple model to obtain a concrete mathematical system that could
be analyzed in detail. It is expected that a wide class of functional forms,
for Eq. \ref{eq:ConcentrationModification}, will still give qualitatively
similar results. In fact, if it is indeed true that such collective
regulatory strategies are used, there are likely to be higher body interactions that would
have evolved to increase its efficacy. This certainly deserves further study.

There is a lot of mathematical similarity between the ideas presented in here, and in many neural network models.
The large number of inputs
to a neuron has suggested that neural networks in the brain employ
different design principles than in normal digital circuits~\cite{LittleShaw,Hopfield,HertzKroghPalmer}.
In these neural network models, computation takes place
collectively. For example, despite only one set of interactions between neurons,
many patterns can be recalled from a
single associative neural network~\cite{LittleShaw,Hopfield}. This is analogous to the ability
of a genetic network with collective regulation to easily adapt to
different rules. The idea here is that the large number of specific interactions that have evolved
between RNA have recorded their evolutionary response to the many separate
environmental conditions. This allows species to call up on
these collective interactions to quickly adapt.

Also as with neural network models, it is always possible to 
implement the same function with more conventional approaches. This
can be done at the expense of adding many more gates in the case
of machine learning, or adding more genes, in the case of genetic
networks. The point is that in both cases, there is another
way of accomplishing the same task, which is considerably more natural and elegant.

The possibility of collective regulation also means that poorly conserved
sequences in the genome
do not imply that they have no useful function. As we have seen
from the simulations,
large changes can be made to the ncRNA involved with collective
regulation, and it can still perform a useful biological function.
Although the hypothesis presented here is speculative, it appears worthwhile to consider
if it could be easily tested experimentally.

\end{document}